\documentclass[a4paper]{jpconf}
\usepackage{graphicx}
\usepackage{lineno}
\def\met{\ensuremath{E_{\mathrm{T}}^{\mathrm{miss}}}}

\def\HT{\ensuremath{H_{\mathrm{T}}}}
\def\ST{\ensuremath{S_{\mathrm{T}}}}
\def\antibar#1{\ensuremath{#1\bar{#1}}}
\def\ttbar{\antibar{t}}
\def\ifb{\mbox{fb$^{-1}$}}

\begin{document}
\title{LHC searches for physics beyond the Standard Model with top quarks}

\author{Tobias Golling, on behalf of the ATLAS and CMS Collaborations}

\address{Department of Physics, Yale University, New Haven, CT, USA}

\ead{tobias.golling@yale.edu}

\begin{abstract}
  Searches are presented for physics beyond the Standard Model
  involving top-quark and related signatures. The results are based on
  proton-proton collision data corresponding to integrated
  luminosities between 1~\ifb\ and 5~\ifb\ collected at a
  center-of-mass energy of 7~TeV with the ATLAS and CMS detectors at
  the Large Hadron Collider in 2011.  The data are found to be
  consistent with the Standard Model.  The non-observation of a signal
  is converted to limits at the 95\% confidence level on the
  production cross section times branching ratio and on the masses of
  the hypothesized new particles for appropriate benchmark models.
\end{abstract}

\section{Introduction}
The Standard Model of particle physics is believed to be an effective
theory valid up to energies close to 1~TeV. However, no new physics
beyond the Standard Model (SM) has been observed yet, and it is
critical to explore a wide range of possible signatures. A promising
avenue lies in final states that involve the heaviest of the particles
presumed to be elementary, the top quark.  This document summarizes
the status of these searches using the ATLAS~\cite{AtlasDetector} and
CMS~\cite{CMSDetector} detectors at the CERN Large Hadron Collider
(LHC).  The results are based on proton-proton collision data
corresponding to integrated luminosities between 1~\ifb\ and 5~\ifb\
collected at a center-of-mass energy of 7~TeV in 2011.

The fact that the top quark is the heaviest elementary particle might
be a hint that it plays a special role in the theory of electroweak
symmetry breaking.  The so-called hierarchy problem refers to the fact
that the large quantum contributions to the square of the Higgs-boson
mass should make the Higgs mass many orders of magnitude larger than
the electroweak scale. Either there is an incredible, unnatural
fine-tuning cancellation or nature chose another mechanism to protect
the Higgs mass to keep it at the observed low
value~\cite{:2012gk,:2012gu}. Many models introduce top partners to
cancel these quantum corrections.  Finally, there are tantalizing
hints of new physics in the forward-backward asymmetry in \ttbar\
events at the Tevatron~\cite{Aaltonen:2011kc,Abazov:2011rq}.

Benchmark signal models are used to define signature-based searches in
final states involving one or two leptons (electrons and muons), jets,
of which typically one or more are required to be identified as
$b$-jets, and missing transverse momentum (\met).  The benchmarks
include models of fourth-generation and vector-like
quarks~\cite{Aad:2012bt,CMS:2012ab,Chatrchyan:2012vu,Aad:2012xc,ATLAS:2012qe,Chatrchyan:2011ay,Chatrchyan:2012yea,ATLAS-CONF-2012-130,Aad:2012bb,CMS-PAS-B2G-12-003,ATLAS:2012aw,Chatrchyan:2012fp,:2012aka,CMS-PAS-EXO-11-066},
top partners in little Higgs models~\cite{Aad:2011wc,Aad:2012uu}, as
well as non-SM production of four top
quarks~\cite{ATLAS-CONF-2012-130}, same-sign top-quark pair ($tt$)
production~\cite{Aad:2012bb}, top+jet resonances in \ttbar+jet
events~\cite{Aad:2012em,Chatrchyan:2012su}, $W' \to t\bar{b}$
resonances~\cite{Aad:2012ej,:2012sc}, and flavor changing neutral
currents (FCNC) in single top-quark production~\cite{Aad:2012gd}.
Searches for \ttbar\ resonances, third generation supersymmetry and
new physics in top-quark decays and properties are covered elsewhere
in these proceedings~\cite{otherTopSearches}.

\section{Experimental techniques}
The experimental approaches are similar between ATLAS and CMS.
Searches are based on the l+jets or dilepton \ttbar\ channels.  Jets
are reconstructed from three-dimensional calorimeter energy clusters
using the anti-$k_t$ jet clustering algorithm~\cite{Cacciari:2008gp}
with a radius parameter of $0.4$ for ATLAS and $0.5$ for CMS. Leptons
are required to pass quality criteria and to be
isolated~\cite{Aad:2011mk,ATLASMuon,CMSEle,CMSMuon}.  The transverse
momenta of jets and leptons are typically required to be larger than
20 or 25~GeV or more.  Multivariate tagging
algorithms~\cite{ATLASBtag,:2012hd} are used to identify $b$-jets.

Typically, the statistical tool BumpHunter~\cite{BumpHunter} is used
to check for deviations, an excess or deficit, from the background
hypothesis. In the absence of a signal, cross-section and mass limits
are derived for benchmark models using CLs~\cite{CLs1,CLs2} or
occasionally Bayesian methods~\cite{Bayesian}.  All limits quoted in
this document are obtained at the 95\% confidence level.  Unless
mentioned otherwise the new particles are assumed to decay with 100\%
branching fraction to the corresponding final state under study.

The largest background typically originates from SM \ttbar\ production
and is estimated from Monte Carlo (MC) simulation using
MC@NLO~\cite{Frixione:2002ik,Frixione:2003ei} or
POWHEG~\cite{Frixione:2007vw} at ATLAS, and
MADGRAPH~\cite{Maltoni:2002qb,Alwall:2007st} at CMS.  Data-driven
multi-jet background estimates are based on the matrix method and on
binned likelihood fits to the \met\ distribution.  Data-driven
$W$+jets estimates use the inherent $W$ charge asymmetry in $pp$
collisions.  ALPGEN~\cite{Mangano:2002ea} (ATLAS) and MADGRAPH (CMS)
are used to model $W$+jets in MC simulation.  The composition of the
flavor of the quarks produced in association with the $W$ boson is
measured in the low jet multiplicity bins and extrapolated using MC
simulations.  Small backgrounds, like the production of single top
quarks, boson pairs, \ttbar+$W$, \ttbar+$Z$, and $Z$+jets are
typically estimated from MC simulations.

The leading systematic uncertainties typically originate from the
uncertainty in the jet energy scale, the $b$-tagging efficiency, and
the \ttbar\ MC modeling.  In certain cases the impact of systematic
uncertainties on the sensitivity of the search is reduced by using
Gaussian constraints or other marginalization techniques, see
e.g. Ref.~\cite{{Aad:2012xc}}.

\section{Results}
An ATLAS analysis~\cite{Aad:2012bt} searches for the pair production
of heavy non-SM quarks $Q$ with decays according to $Q\bar{Q} \to
W^+qW^-\bar{q}$ with $q = d$, $s$, $b$ for up-type $Q$ or $q = u$, $c$
for down-type $Q$.  The search is performed with 1.04~\ifb\ of
integrated luminosity. Dilepton final states are selected, requiring
large \met\ and at least two jets. No $b$-tagging is applied.  Mass
reconstruction of heavy quark candidates is performed by assuming that
the $W$-boson decay products are nearly collinear. The resulting mass
reconstruction is shown in Fig.~\ref{ATLAS_QQdilep}.  No deviation
from the SM expectation is observed.  Heavy non-SM quark masses below
350~GeV are excluded.

\begin{figure}[h]
 \begin{minipage}{17.9pc}
 \includegraphics[width=17.9pc]{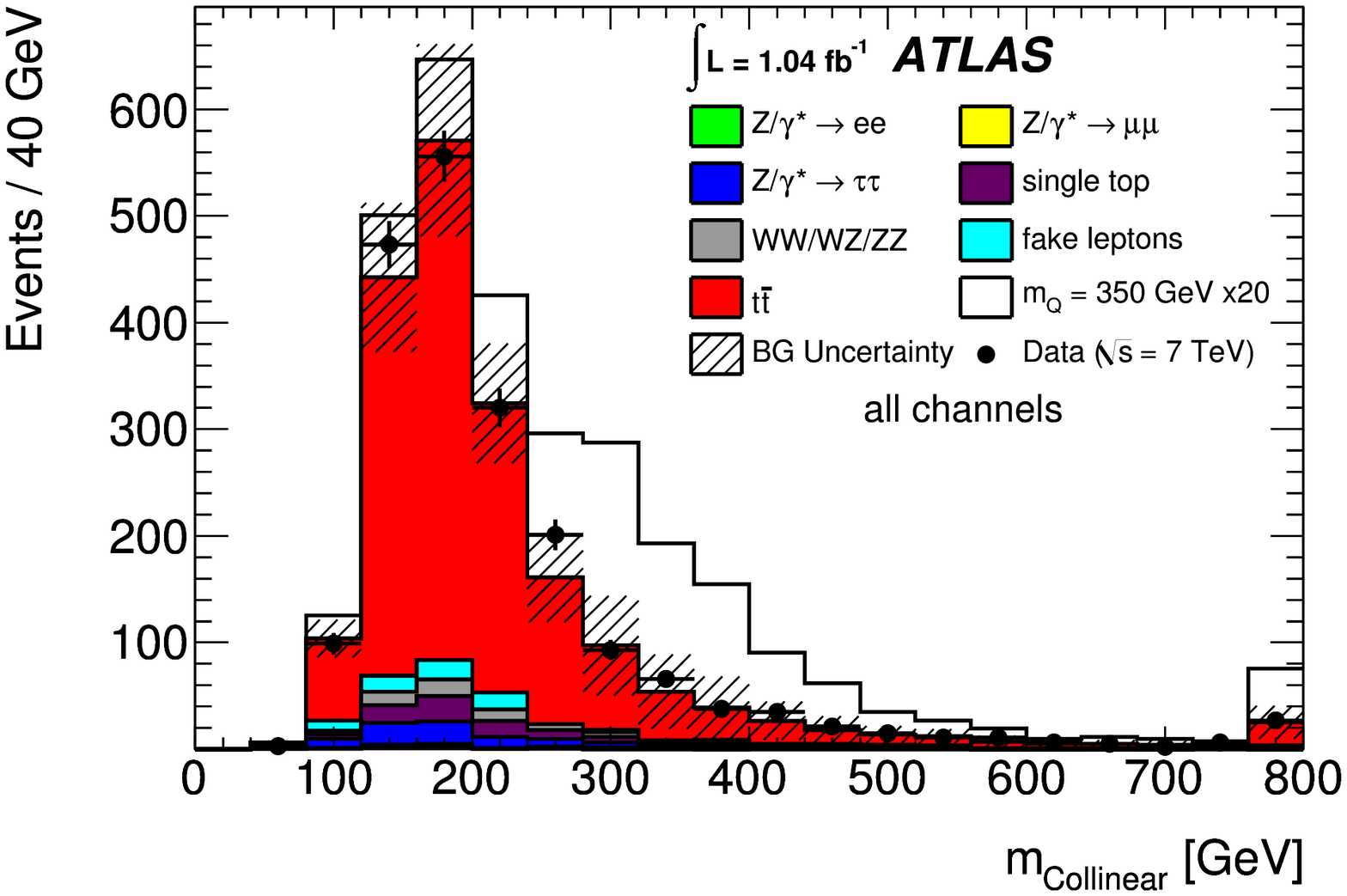}
 \caption{\label{ATLAS_QQdilep}Expected and observed distributions of
   the collinear mass for the sum of $ee$, $\mu\mu$ and $e\mu$
   channels~\cite{Aad:2012bt}. The last bin contains overflow
   events. Samples are stacked.  The signal has been amplified to 20
   times the expected rate. }
 \end{minipage}\hspace{2pc}%
 \begin{minipage}{17.9pc}
 \includegraphics[width=17.9pc]{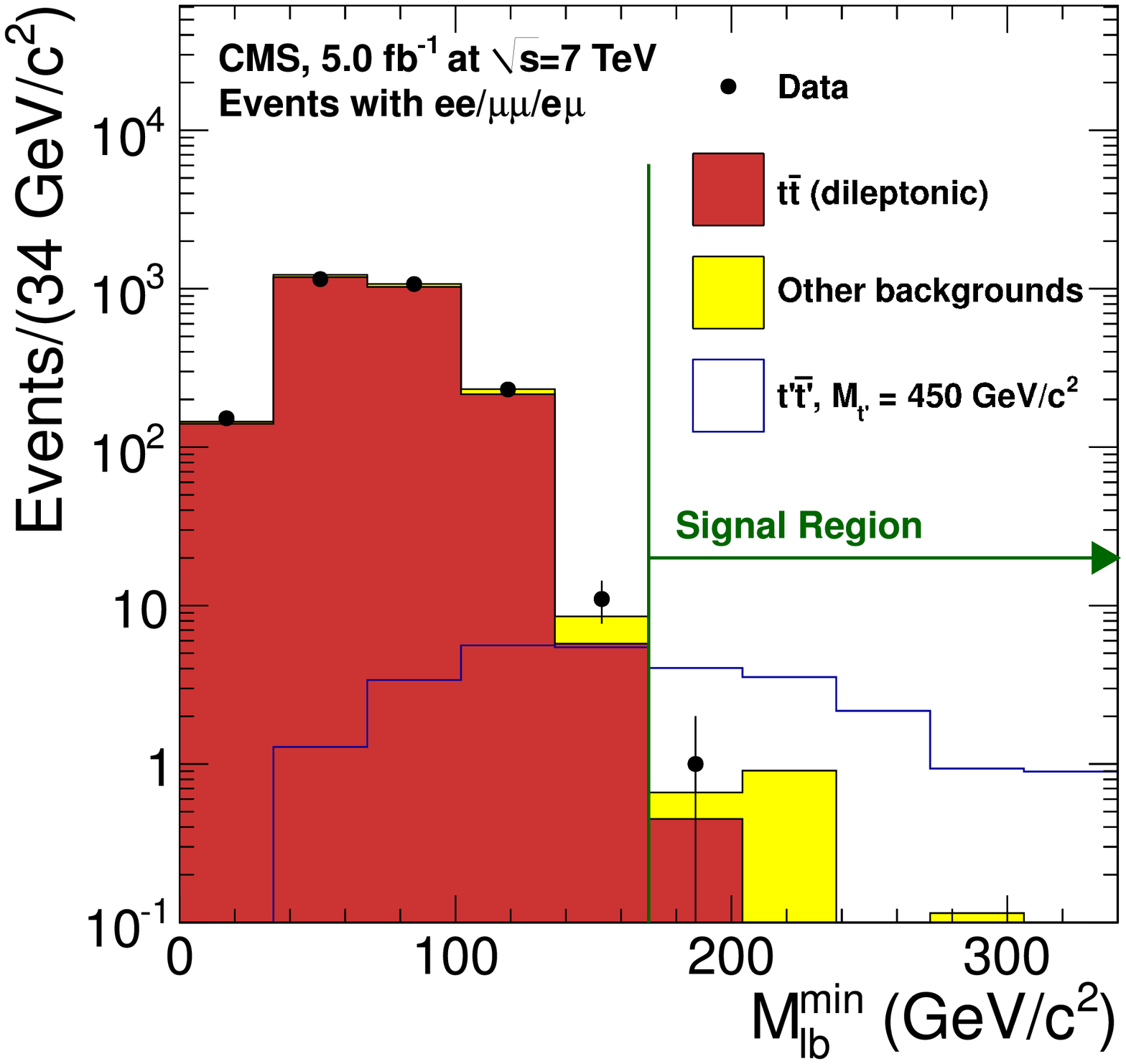}
 \caption{\label{CMS_Mlbmin}Comparison between the data and the
   simulated background for
   $M_{lb}^{\mathrm{min}}$~\cite{CMS:2012ab}. The expected
   distribution for a signal with $m_{t'}=450$~GeV is also shown.  The
   signal region is defined by $M_{lb}^{\mathrm{min}} >170$~GeV.}
 \end{minipage} 
\end{figure}

A similar CMS search~\cite{CMS:2012ab} for pair production of heavy
top-like quarks $t'$ has been performed in the decay mode $t' \bar{t}'
\to W^+bW^-\bar{b}$. The search uses 5.0~\ifb\ of integrated
luminosity.  Again dilepton final states are selected, requiring large
\met\ and at least two jets, but this time exactly two of the jets
have to be identified as $b$-jets.  The minimum value of the four
possible masses of the system defined by one of the two leptons and
one of the two $b$-jets ($M_{lb}^{\mathrm{min}}$) is found to be a
good variable for distinguishing $t' \bar{t}'$ from $t\bar{t}$ events,
as can be seen in Fig.~\ref{CMS_Mlbmin}.  The observed number of
events agrees with the expectation from SM processes.  Heavy $t'$
quarks with a mass less than 557~GeV are excluded.

Searches for heavy top-like quarks $t'$ given the hypothesized decay
mode $t' \bar{t}' \to W^+bW^-\bar{b}$ are also conducted in final
states with a single charged lepton, \met\ and, depending on the
analysis, at least three or four jets, of which at least one must be
identified as a
$b$-jet~\cite{Chatrchyan:2012vu,Aad:2012xc,ATLAS:2012qe}.

CMS~\cite{Chatrchyan:2012vu} uses the reconstructed $t'$ mass
$m_{\mathrm{reco}}$, obtained from a kinematic fit of the
reconstructed four-momenta to the decay hypothesis $t' \bar{t}' \to
W^+bW^-\bar{b} \to l\nu bq\bar{q}'\bar{b}$, as well as \HT, defined as
the scalar sum of the transverse momenta of the objects associated to
the $t'$ and $\bar{t}'$ decay products. The two-dimensional
distributions of \HT\ versus $m_{\mathrm{reco}}$ are fitted with
analytic functions for the signal S and the background B. All
two-dimensional bins are then sorted in increasing order of the
expected S/B ratio, using the functions.  The resulting distribution
of this so-called S/B rank is shown in Fig.~\ref{CMS_SoBrank} and is
used to exclude $t'$ masses below 570~GeV with 5.0~\ifb\ of integrated
luminosity.

\begin{figure}[h]
 \begin{minipage}{17.9pc}
 \includegraphics[width=17.9pc]{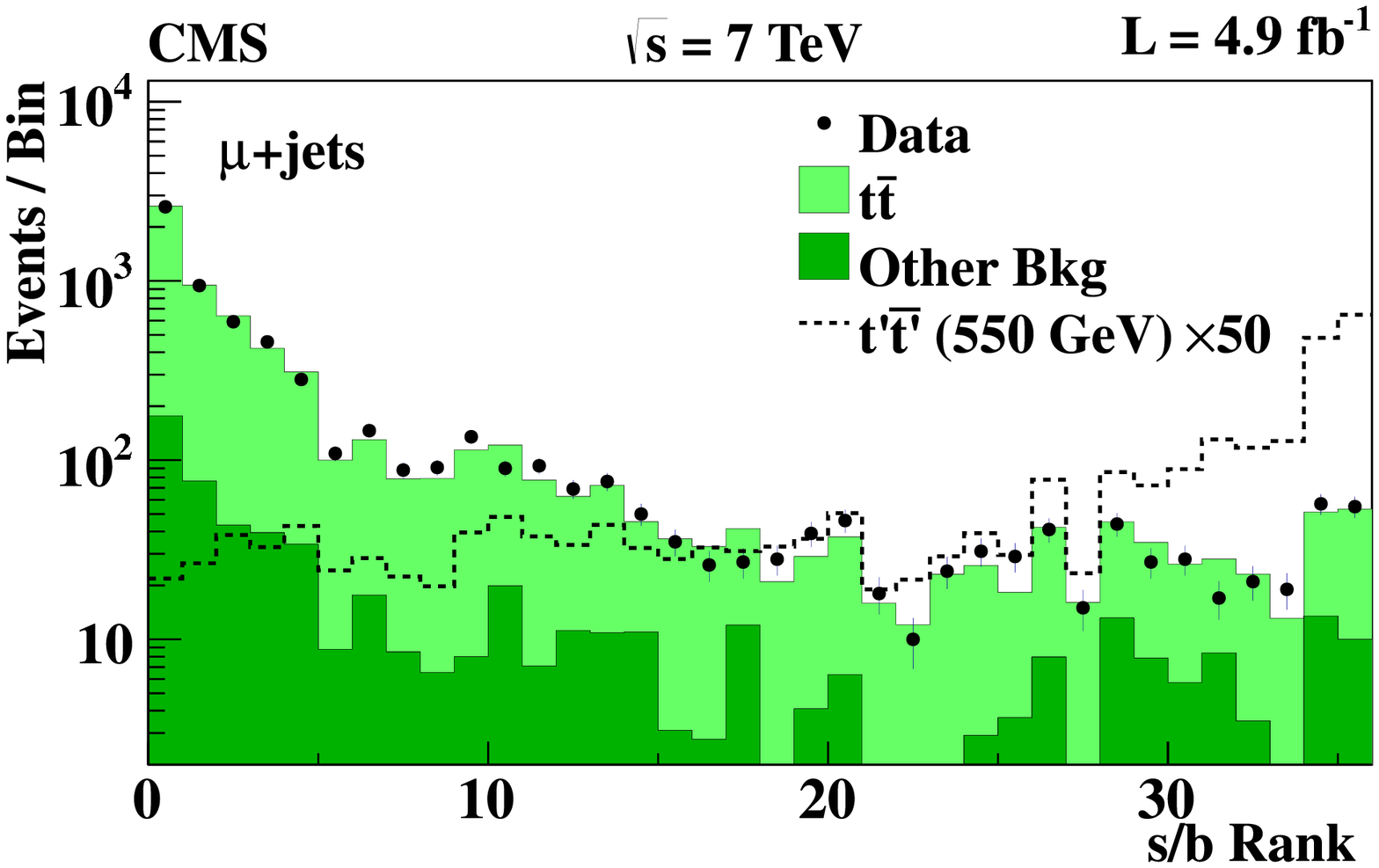}
 \caption{\label{CMS_SoBrank}Number of events per bin in the
   two-dimensional \HT\ versus $m_{\mathrm{reco}}$ histogram, as a
   function of the S/B rank for the $\mu$+jets
   channel~\cite{Chatrchyan:2012vu}.}
 \end{minipage}\hspace{2pc}%
 \begin{minipage}{17.9pc}
  \includegraphics[width=17.9pc]{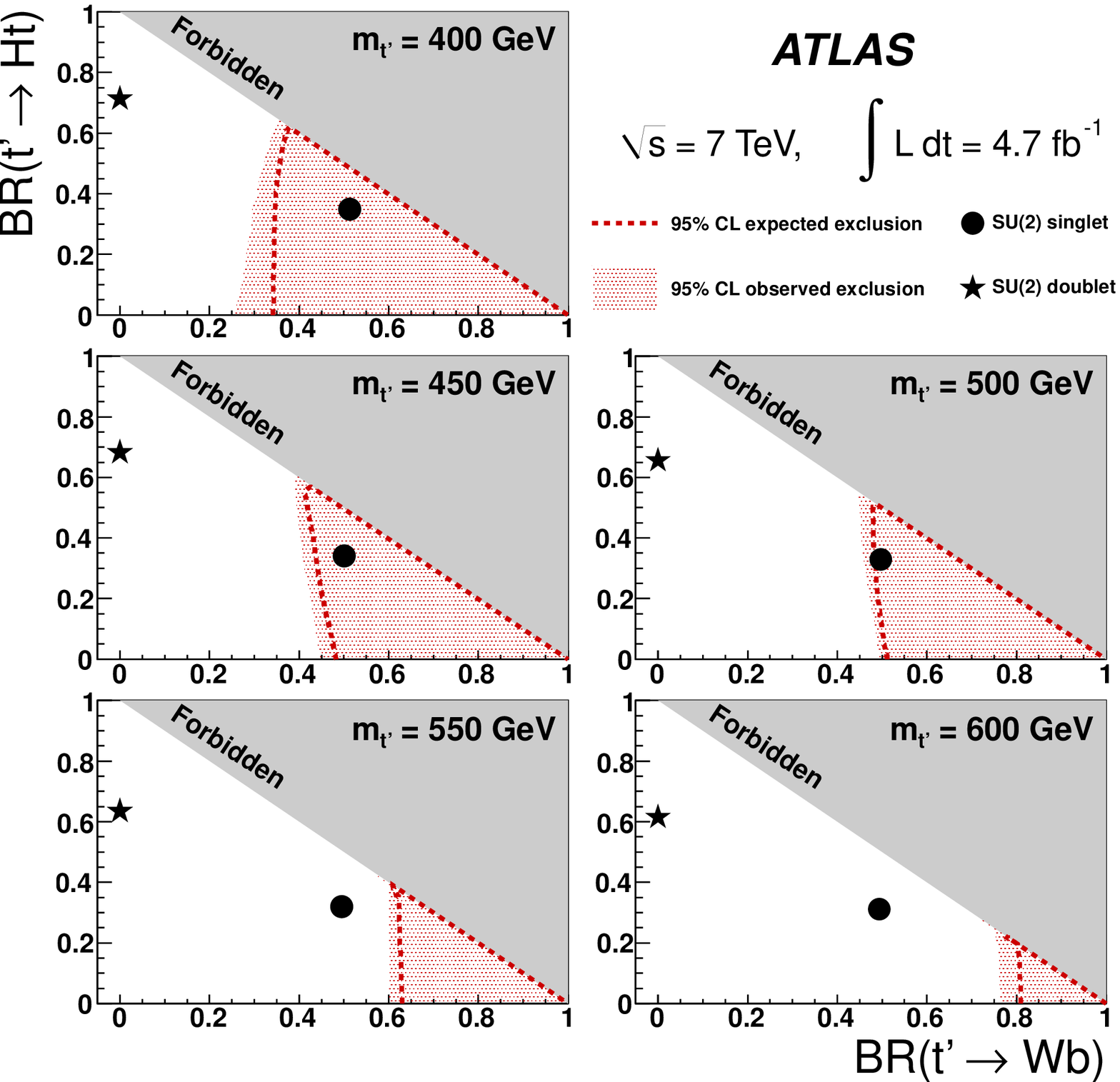}
  \caption{\label{ATLAS_VLQplane}Observed (red filled area) and
    expected (red dashed line) exclusion limits in the plane of
    BR($t'\to Wb$) vs BR($t'\to Ht$) for different values
    of the vector-like $t'$ quark mass~\cite{ATLAS:2012qe}. The grey
    solid area corresponds to the unphysical region where the sum of
    the branching ratios exceeds unity.}
 \end{minipage} 
\end{figure}

ATLAS~\cite{ATLAS:2012qe} uses the reconstructed mass
$m_{\mathrm{reco}}$ of the candidate $t'$ as the discriminant.  A
tight selection is applied targeting $t'$ masses above 400~GeV.  In
this mass range the decay products have large momenta.  The
reconstruction of hadronically-decaying $W$ bosons $W_{\mathrm{had}}$
takes advantage of this.  They are either defined as a single jet with
$p_T > 250$~GeV and jet mass in the range of 60-110~GeV or as a dijet
system with $p_T > 150$~GeV, angular separation $\Delta R(j, j) <
0.8$, and mass within the range of 60-110~GeV.  Additional kinematic
selection criteria include \HT\ (defined as the scalar sum of the
transverse momenta of the lepton, \met, and the jets from the
hypothesized $t'$ decays) to be larger than 750~GeV, $\Delta R(l, \nu)
< 1.4$, $\Delta R(W_{\mathrm{had}}, b$-jet$) > 1.4$, and $\Delta R(l,
b$-jet$) > 1.4$.  With 4.7~\ifb\ of integrated luminosity a $t'$ quark
with mass lower than 656~GeV is excluded.

In addition, in light of the recent discovery of a new boson of mass
126~GeV at the LHC, upper limits are derived, as shown in
Fig.~\ref{ATLAS_VLQplane}, for vector-like quarks of various masses in
the two-dimensional plane of BR($t'\to Wb$) versus BR($t'\to Ht$),
where $H$ is the SM Higgs boson (BR($t'\to Zt$) $=1-$ BR($t'\to Wb$)
$-$ BR($t'\to Ht$).)

A search for pair-produced, heavy, vector-like charge-$2/3$ quarks is
performed by CMS~\cite{Chatrchyan:2011ay}, assuming BR($t'\to Zt$)$=1$.
Events are selected by requiring two charged leptons from the
$Z$-boson decay, as well as an additional isolated charged
lepton. Using 1.14~\ifb\ of integrated luminosity $t'$ quarks with a
mass less than 475~GeV are excluded.

Both ATLAS and CMS conduct searches for heavy pair-produced
bottom-like quarks.  These $b'$ quarks are assumed to decay
exclusively to $Wt$.  The $W^+tW^-\bar{t}$ final state has the
distinctive signature of three or more leptons or two leptons of same
charge which is exploited.

CMS uses 4.9~\ifb\ of integrated luminosity to select trilepton and
same-sign-dilepton events with \met~\cite{Chatrchyan:2012yea}.  At
least one jet must be identified as a $b$-jet.  Events are rejected
where any two leptons of the same flavor have an invariant mass
consistent with the $Z$-boson mass.  Furthermore, \ST, defined as the
scalar sum of the transverse momenta of the leptons, \met, and the
jets, is used to reject background, as shown in Fig.~\ref{CMS_trilep}.
$b'$ quarks with mass below 611~GeV are excluded.

\begin{figure}[h]
 \begin{minipage}{17.9pc}
\includegraphics[width=17.9pc]{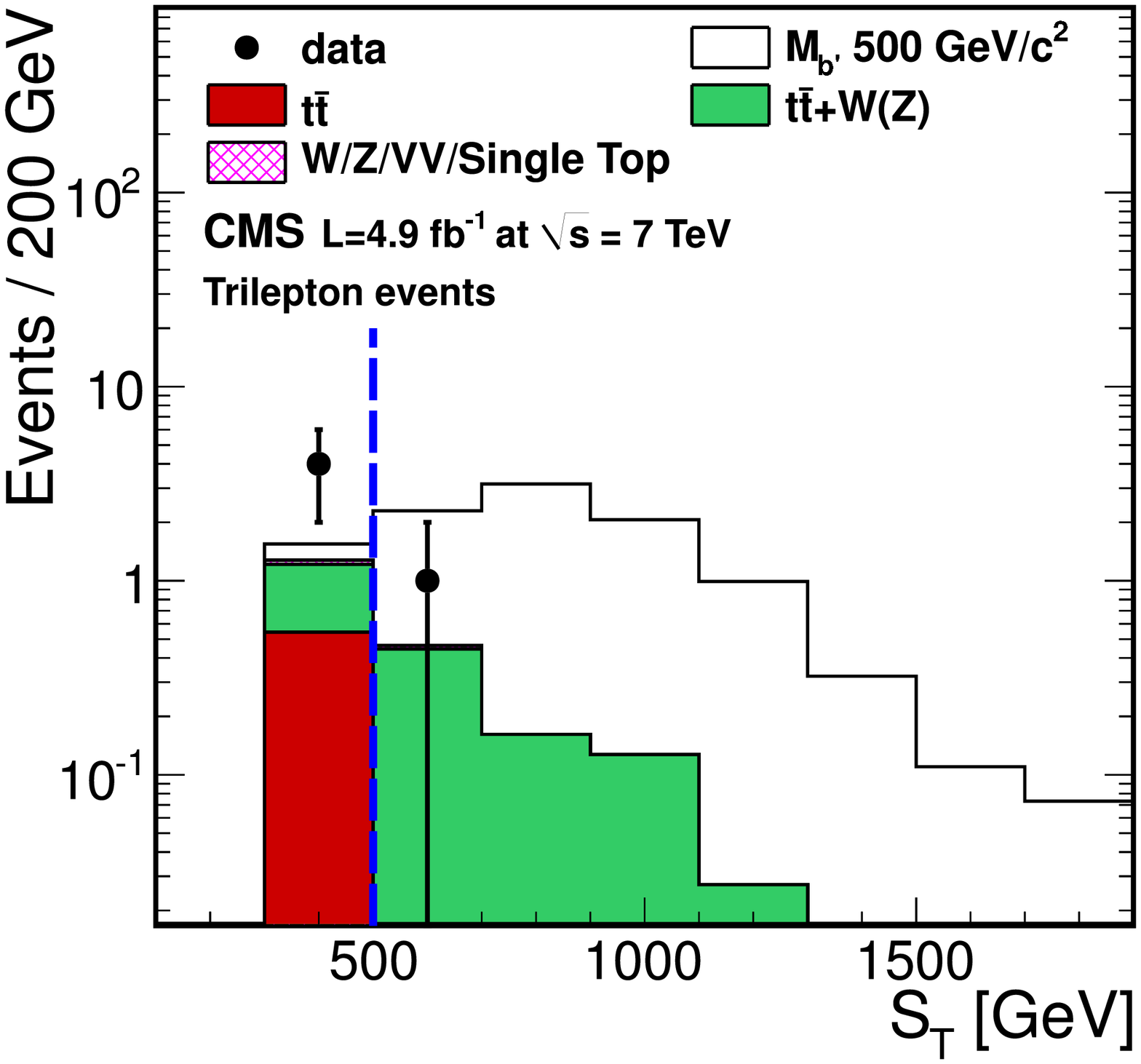}
\caption{\label{CMS_trilep}The \ST\ distributions for the trilepton
  channel~\cite{Chatrchyan:2012yea}.  The vertical dotted line
  indicates the lower \ST\ threshold used in the analysis.}
 \end{minipage}\hspace{2pc}%
 \begin{minipage}{17.9pc}
\includegraphics[width=17.9pc]{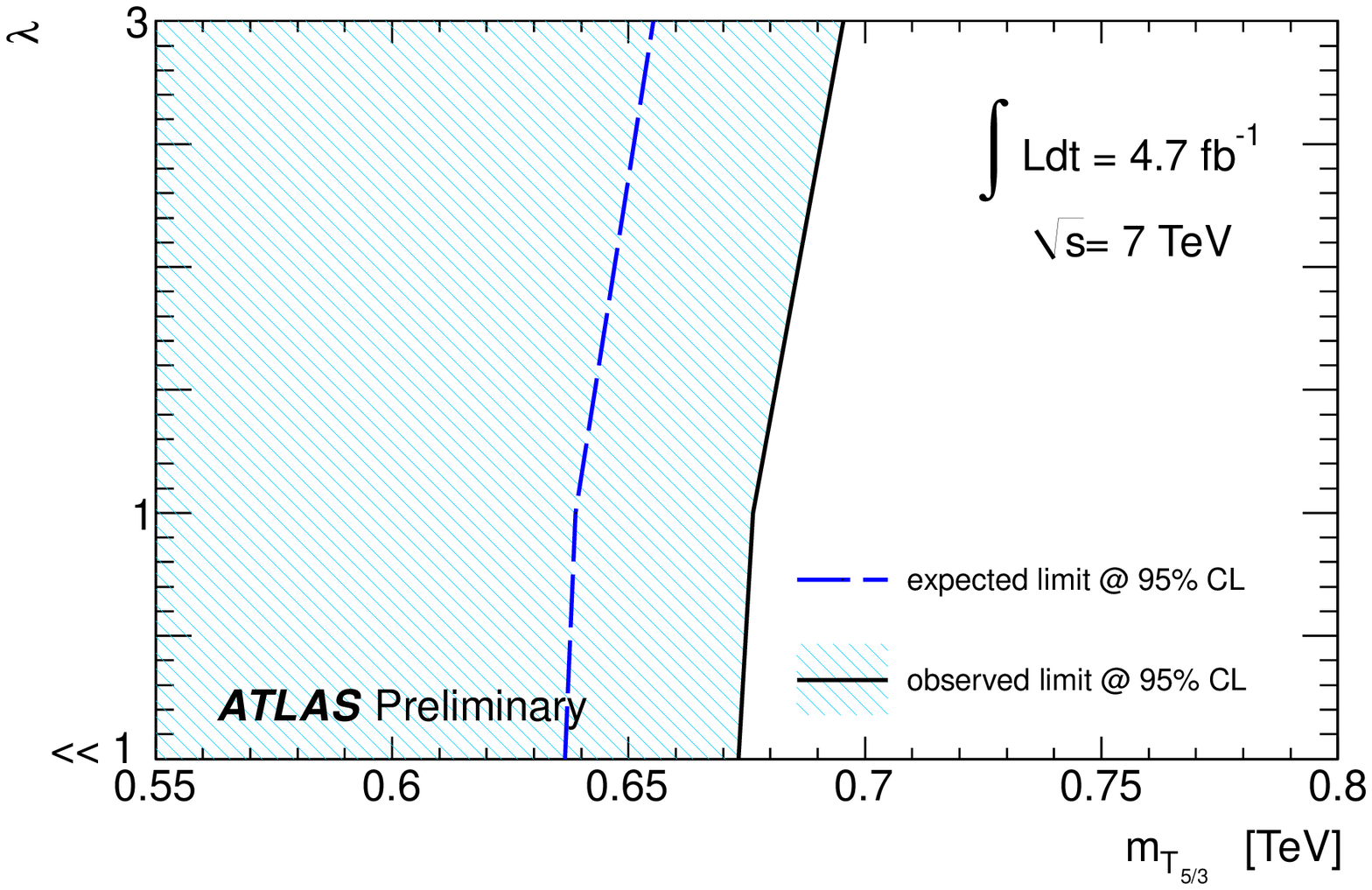}
\caption{\label{ATLAS_T5/3}Expected and observed lower limits on the
  $T_{5/3}$ signal as a function of the $T_{5/3}$ mass and the
  coupling constant $\lambda$~\cite{ATLAS-CONF-2012-130}. The shaded
  area is excluded.}
 \end{minipage} 
\end{figure}

A similar analysis by ATLAS focuses on same-sign-dilepton events using
4.7~\ifb\ of integrated luminosity~\cite{ATLAS-CONF-2012-130}.  The
most important background is the contribution arising from fake
leptons. Another significant background is from various SM processes
where two real leptons are produced, but in which one of the leptons
has a mis-identified charge.  Additional signal hypotheses are
considered for the interpretation of the results.  Both single and
pair production of new heavy quarks $T_{5/3}$, with charge $5/3$, are
considered with $T_{5/3}\to W^+t$.  For the single production the
assumed coupling constant $\lambda$ of the $tWT_{5/3}$ vertex is
varied as shown in Fig.~\ref{ATLAS_T5/3}.  Assuming only $b'\bar{b}'$
or $T_{5/3}\bar{T}_{5/3}$ production quark masses below 670~GeV are
excluded.  In addition, first limits are set on non-SM production of
four top quarks, yielding $\sigma_{\mathrm{4tops}}<61$~fb.
In an earlier version of the ATLAS same-sign-dilepton
search~\cite{Aad:2012bb} using 1.04~\ifb\ of integrated luminosity the
results are also interpreted for same-sign top-quark pair production.  The
results leave little room to explain the measurement of the
forward-backward asymmetry in top-quark pair production at the
Tevatron by a flavor-changing $Z'$ boson.

CMS excludes $T_{5/3}$ masses below 645~GeV by analyzing very similar
same-sign-dilepton final states, assuming $T_{5/3}\bar{T}_{5/3}$
production and using 5.0~\ifb\ of integrated
luminosity~\cite{CMS-PAS-B2G-12-003}.

Also the 1-lepton channel is used by ATLAS to set limits on $b'\bar{b}'$
production with $b' \to W^+t$~\cite{ATLAS:2012aw}.  Similar to
Ref.~\cite{ATLAS:2012qe} hadronically decaying $W$ bosons are
reconstructed.  The limits are not competitive since only 1.04~\ifb\
of integrated luminosity was used for these results.

A combined search in the 1-lepton, same-sign-dilepton and trilepton
final states is presented by CMS by assuming both single and pair
production of both $t'$ and $b'$~\cite{Chatrchyan:2012fp}.  By
analyzing \ST\ and the invariant $bW$ mass in bins of number of
reconstructed hadronically decaying $W$ bosons and $b$-jets, limits
are presented as a function of the mass difference between $t'$ and
$b'$ and as a function of the matrix element $V_{tb}$.

Both ATLAS~\cite{:2012aka} and CMS~\cite{CMS-PAS-EXO-11-066} set
limits on $b'\bar{b}'$ production with at least one $b'$ decaying to a
$Z$ boson and a bottom quark.  Using 2.0~\ifb\ of integrated
luminosity ATLAS excludes vector-like singlet $b'$ quarks mixing
solely with the third SM generation with masses below 358~GeV.
Assuming BR($b'\to Zb$)$=1$ CMS excludes $b'$ masses below 550~GeV
using 4.9~\ifb.

Searches for signatures of pair production of supersymmetric top
partners also have sensitivity to spin-$1/2$ top partners in little
Higgs models as shown by ATLAS analyses presented in
Refs.~\cite{Aad:2011wc,Aad:2012uu}.

Searches for new heavy resonances, a color singlet $W'$ or a color
triplet $\phi$, produced in association with a top quark are motivated
by top-flavor violating processes designed to explain the $t\bar{t}$
forward-backward asymmetry observed at the Tevatron.  Two-dimensional
limits are set on the mass and the coupling of $W'$ and $\phi$ by
ATLAS~\cite{Aad:2012em} by analyzing the $\bar{t}$+jet and the $t$+jet
invariant mass, respectively, in $t\bar{t}$+jet candidate events.
Similar limits are set by CMS~\cite{Chatrchyan:2012su}.  The limits
leave little room for top-flavor violating processes to explain the
$t\bar{t}$ forward-backward asymmetry observed at the Tevatron.

Both ATLAS~\cite{Aad:2012ej} and CMS~\cite{:2012sc} search for
resonances in the $t\bar{b}$ (and c.c.) spectrum and set lower limits
on the mass of right handed $W' \to t\bar{b}$ of 1.85~TeV. ATLAS uses
the invariant $t\bar{b}$ mass as the discriminant, while CMS makes use
of boosted decision trees.

ATLAS uses a neural network analysis to search for FCNC single
top-quark production~\cite{Aad:2012gd}. Two-dimensional limits are set
in the plane of BR($t \to ug$) and BR($t \to cg$).

\section{Conclusions}
Top quarks play an important role in ATLAS and CMS searches for
physics beyond the SM.  No hints of new phenomena could be established
yet.  Results from LHC $pp$ collisions at a center-of-mass energy of
8~TeV and eventually close to 14~TeV are anticipated with great
suspense.

\ack{The author is supported by grants from the Department of Energy
  Office of Science and by the Alfred P. Sloan Foundation.}

\end{document}